# Efficient Pixelated Rectenna using Binary Optimization for WPT Applications

Rasool Keshavarz[*1], Md. Amanath Ullah[*#2], Ali Raza[*#3], and Negin Shariati[*#4]

[*]RF and Communication Technologies (RFCT) Research Laboratory, School of Electrical and Data Engineering, Faculty of Engineering and IT, University of Technology Sydney, Ultimo, NSW 2007, Australia.

[#]Food Agility CRC Ltd, Sydney, NSW, Australia 2000.

{[1]Rasool.Keshavarz, [4]Negin.Shariati}@uts.edu.au, {[2]Mdamanath.Ullah, [3]Ali.Raza-1}@student.uts.edu.au

***Abstract* — This paper introduces a highly efficient rectenna (rectifying antenna) using a binary optimization algorithm. A novel pixelated receiving antenna has been developed to match the diode impedance of a rectifier, eliminating the need for a separate matching circuit in the rectenna's rectifier. The receiving antenna configuration is fine-tuned via a binary optimization algorithm. A rectenna is designed using optimization algorithm at 2.5 GHz with 38% and 67% RF-DC conversion efficiency when subjected to 0 dBm and 10 dBm incident power with output voltages of 815mV and 2.49 V, respectively. The proposed rectenna demonstrates versatility across various low-power WPT (wireless power transfer) applications.**

## I. INTRODUCTION

Energy-autonomous or self-powered devices are poised to become indispensable components within future wireless sensor networks (WSNs) and the Internet of Things (IoT). [1, 2]. In remote or inaccessible regions, the primary challenge with battery-powered electronic devices and sensors lies in the need for battery replacement or maintenance, which demands physical access and upkeep. Wireless Power Transfer (WPT) has gained widespread adoption across diverse applications as depicted in Fig. 1, addressing prevalent battery-related challenges. Ambient Radio Frequency (RF) energy harvesting (RFEH) and WPT stand out as highly promising solutions for addressing power supply challenges encountered by low-power devices and sensors [3-5]. Over the past decade, there has been considerable interest in wireless power transfer technology, particularly when paired with RFEH. This interest is fueled by several factors: the capability to transmit and receive wireless power over long distances (Far-field WPT) [6], the penetration of RF signals into various structures such as walls, bridges, and tunnels, the potential for harvesting RF energy throughout the day, the on-demand availability of power through dedicated RF power sources (e.g., WPT), and the growing utilization of IoT devices, wireless sensor nodes, and low-power electronics [7-10].

One of the pivotal components in wireless power transmission systems is the rectenna, responsible for receiving RF waves and converting them into DC power. A rectenna consists primarily of two components: a receiving antenna and a rectifier, both crucial for the process of capturing RF energy [11]. While patch-type rectennas have been extensively studied [12-15], prior research often employed conventional patch antenna designs for the receiving antenna.

Optimization algorithms play a crucial role in antenna design, particularly when balancing factors such as gain, operating frequency, input impedance and size constraints, which are challenging to address using conventional equations alone. By employing optimization algorithms, antenna designers can effectively navigate this intricate design landscape, enabling the creation of highly efficient and compact antennas tailored to specific requirements across various applications [16].

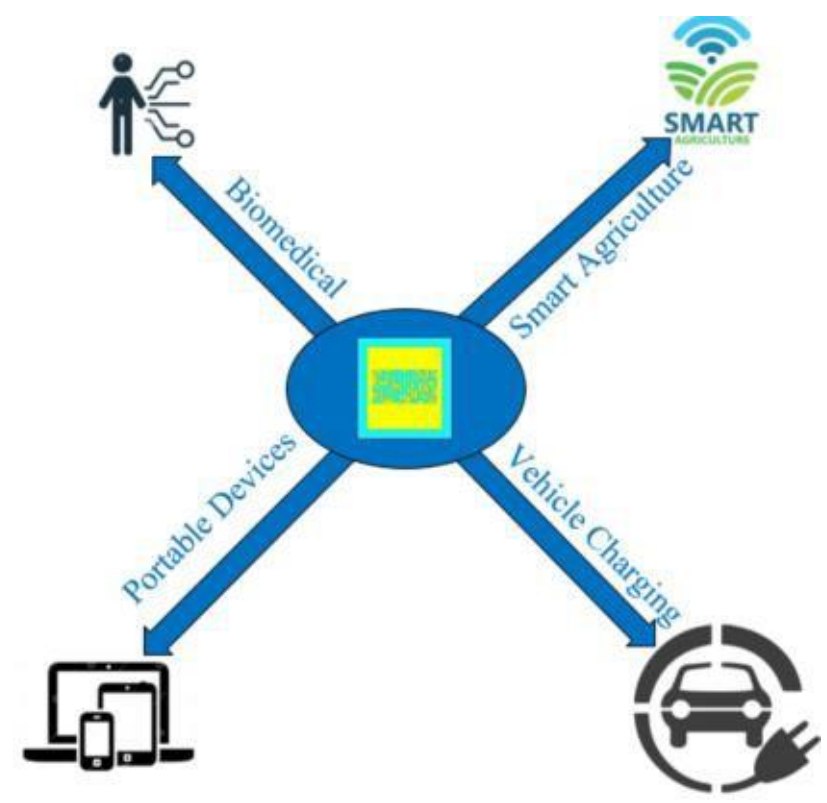

Fig.1. Various application of WPT.

Intelligent optimization algorithms have supplanted traditional electromagnetic simulator optimization techniques in antenna design due to their superior effectiveness in achieving desired outcomes. Optimization algorithms enable the exploration of numerous alternative geometric configurations to design viable structures and meet design constraints. An exemplary instance is Particle Swarm Optimization (PSO) [16-18]. The core PSO approach encounters challenges related to premature convergence and struggles with high-dimensional or multi-objective problems. Researchers have been actively enhancing the performance of standard PSO [19, 20] to mitigate these issues. Additionally, the application of PSO can be extended to antenna designs with discrete shapes using binary PSO (BPSO). Pixelated antenna design offers flexibility in achieving diverse design objectives, such as single or multiband compact antenna design with enhanced BPSO [20, 21]. The utilization of novel patch shapes for receiving antennas can significantly enhance RF energy harvesting or WPT capabilities, as they are not constrained by the conventional length and width limitations of patch rectennas.

Matching the antenna impedance with the rectifier impedance necessitates a matching network. However, designing such a circuit for a rectenna poses considerable challenges. Furthermore, the intricate structure of the matching network can potentially reduce efficiency and escalate fabrication costs. Conventional matching networks involve circuit components that introduce additional complexity into the design and can detrimentally affect overall efficiency [22, 23]. Therefore, there is a preference to maintain overall efficiency and reduce design complexities by eliminating the physical matching network. Furthermore, avoiding the use of a matching circuit could result in a more compact size for the rectenna.

(a)

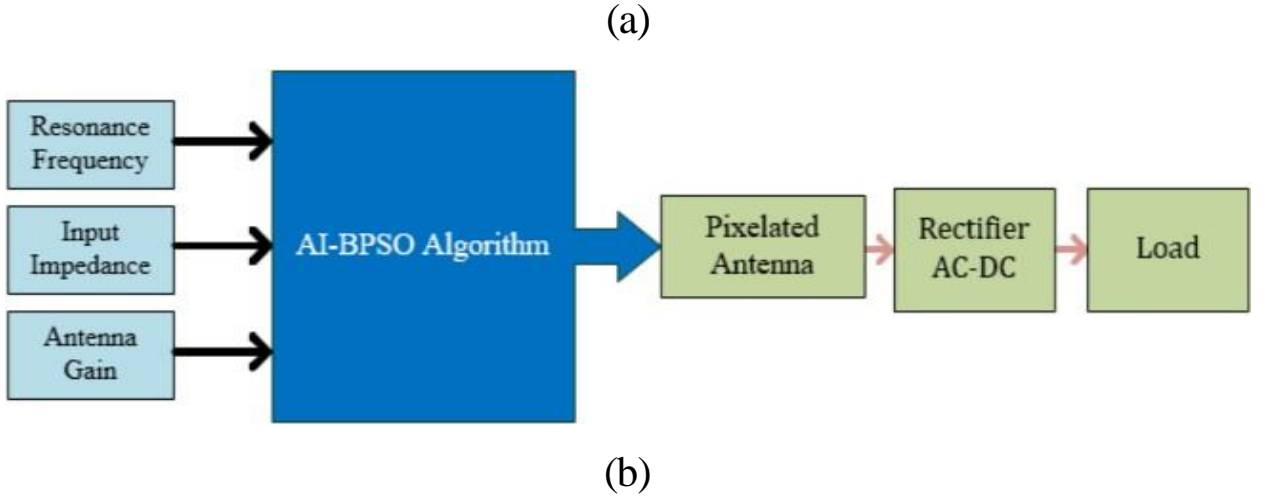

(b)

Fig. 2. (a) Conventional rectenna configuration with matching network; (b) proposed rectenna design without external matching network.

An experimental setup, illustrated in Fig. 11, is employed to assess the performance of the proposed rectenna. Utilizing a vector signal generator (R&S SMW200A) as the RF power source, supplemented by a ZHL-15W-422 amplifier to mitigate path loss, ensures accurate power at the input of the antenna. A log-periodic antenna with 4.5 dBi gain serves as the transmitter, while the rectenna is positioned at the receiver, maintaining a distance of 30 cm. Fig. 12 showcases the results of the proposed rectenna's output DC voltage, RF-DC conversion efficiency, and output DC power plotted against input RF power at 2.5 GHz. An efficiency of 37% is attained at a low power level of 0 dBm, with the maximum efficiency peaking at 64% at +12 dBm.

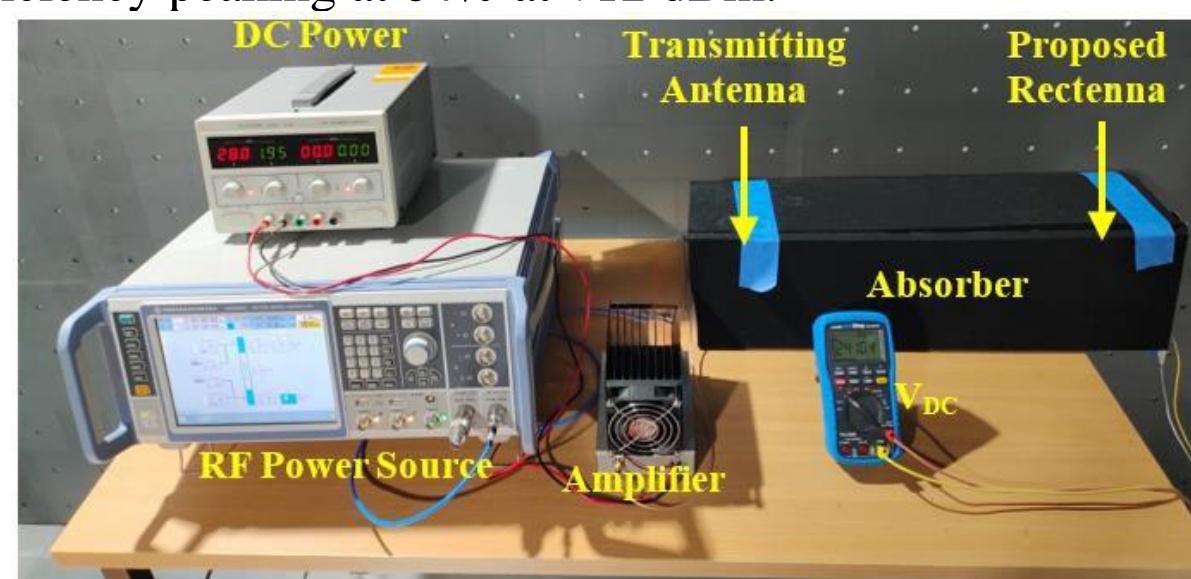

Fig.. Measurement setup to evaluate the performance of the proposed rectenna.

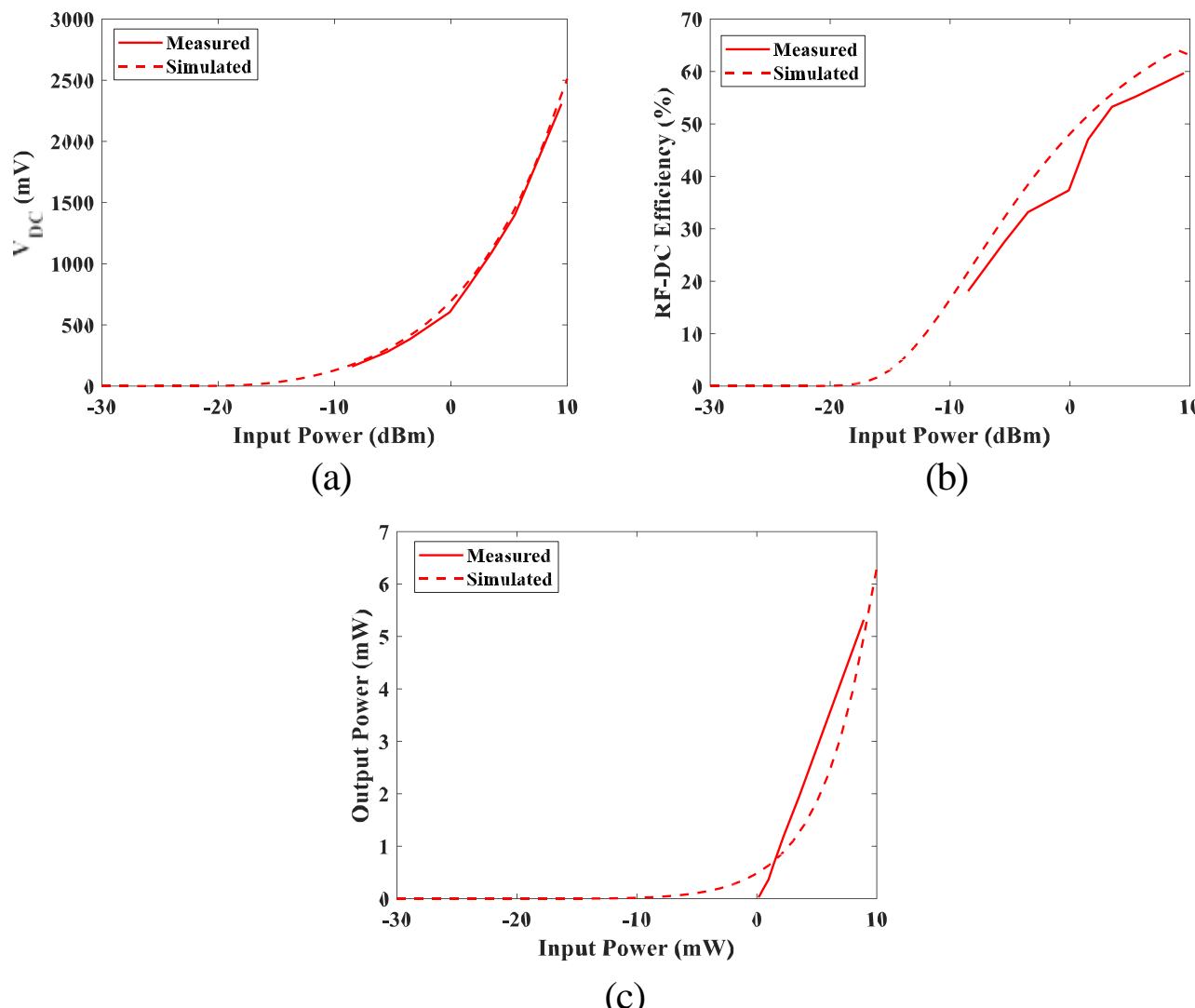

Fig. 12. Performance of the proposed rectenna, (a) output dc voltage, (b) RF-DC conversion efficiency, (c) output power vs input power.

## I. Conclusion

This paper introduces an efficient design of a pixelated rectenna. The proposed rectenna comprises a pixelated receiving antenna that has been meticulously optimized to match the complex impedance of the rectifier diode, thus obviating the necessity for a separate matching circuit. The optimization process is facilitated by a BPSO algorithm. Achieving an efficiency of 37% at a low power level of 0 dBm and 64% at +12 dBm, the proposed rectenna emerges as a promising candidate for RFEH and WPT applications in low-power devices.